\documentclass[pre,showpacs,twocolumn,longbibliography]{revtex4-1}

\usepackage{hyperref}
\usepackage{color}
\usepackage[usenames,dvipsnames]{xcolor}
\usepackage{amsmath,amsthm,amssymb}
\usepackage{graphicx}
\usepackage{epsfig}
\usepackage{dcolumn}
\usepackage{bm}
\usepackage{mathrsfs}
\usepackage{multirow}
\usepackage[all]{xy}
\usepackage{pbox}
\usepackage{verbatim}
\usepackage[latin1]{inputenc}
\usepackage{braket}
\usepackage{mathtools}
\usepackage{bm}
\usepackage{tikz}
\usepackage{xcolor}
 
\usepackage{mathtools}

\usepackage{mathtools}

\def\(({\left(}
\def\)){\right)}
\def\[[{\left[}
\def\]]{\right]}

\newcommand{\be}{\begin{equation}}
\newcommand{\ee}{\end{equation}}
\newcommand{\ben}{\begin{eqnarray}}
\newcommand{\een}{\end{eqnarray}}
\newcommand{\beq}{\begin{equation}}
\newcommand{\eeq}{\end{equation}}

\newcommand{\xddots}{%
  \raise 4pt \hbox {.}
  \mkern 6mu
  \raise 1pt \hbox {.}
  \mkern 6mu
  \raise -2pt \hbox {.}
}

\begin{document}  

\title{Trajectory phase transitions in non-interacting spin systems }

\author{Loredana M. Vasiloiu}
\author{Tom H. E. Oakes}
\author{Federico Carollo}
\author{Juan P. Garrahan}
\affiliation{School of Physics and Astronomy}
\affiliation{Centre for the Mathematics and Theoretical Physics of Quantum Non-Equilibrium Systems,
University of Nottingham, Nottingham, NG7 2RD, UK}

\date{\today}

\begin{abstract}

We show that a collection of independent Ising spins evolving stochastically can display surprisingly large fluctuations
towards ordered behaviour, as quantified by certain types of time-integrated {\em plaquette observables}, despite the 
underlying dynamics being non-interacting.
In the large deviation (LD) regime of long times and large system size, this can give rise to a phase transition in trajectory space. 
As a non-interacting system we consider a collection of spins undergoing single spin-flip dynamics at infinite-temperature. For the dynamical observables we study, the associated tilted generators have an exact and explicit spin-plaquette duality.
Such setup suggests the existence of a transition (in the large size limit) at the self-dual point of the tilted generator. The nature of the LD transition depends on the observable. We consider explicitly two situations: (i) for a pairwise bond observable the LD transition is continuous, and equivalent to that of the transverse field Ising model; (ii) for a higher order plaquette observable, in contrast, the LD transition is first order. Case (i) is easy to prove analytically, while we confirm case (ii) numerically via an efficient trajectory sampling scheme that exploits the non-interacting nature of the original dynamics.
\end{abstract}

\maketitle 

\section{Introduction}
Phase transitions \cite{Chandler1987,Binney,Zinn-Justin,Sachdev,Perez} occur when a physical system undergoes a sudden structural change as a reaction to an infinitesimal variation of a suitable control parameter across a critical value. The abrupt modification of the macroscopic properties of the system is reflected in a non-analytic behaviour of an order parameter \cite{Chandler1987,Binney,Zinn-Justin}. In classical equilibrium the parameter driving the phase transition is an intensive field conjugate to the order parameter, such as the inverse temperature or a chemical potential, while in quantum phase transitions it is a coupling constant in the Hamiltonian, such as the strength of an external field or of the interactions. 
More recently, the notion of phase transition has been extended to include also critical phenomena taking place in large fluctuations of nonequilibrium processes, e.g.\ \cite{Bertini,Bertini2,Derrida,Dammer,BLYTHE,Merolle2005,Garrahan2007,Lecomte,Garrahan2009,Hedges2009,Jack,Speck2012,Jack2015} (see \cite{TOUCHETTE,touchette2011,Garrahan2018,Jack2019} for reviews).
In this scenario, sudden changes in the spatio-temporal structures of the system (trajectories) are witnessed by a non-analytic behaviour in free-energy or entropy-like functionals describing the statistics of an appropriate time-integrated observable. 

Similarly to what happens in equilibrium statistical mechanics, where the emergence of a phase transition requires the presence of interactions and of an infinite number of degrees of freedom, one would expect to observe these large deviation (dynamical) phase transitions uniquely in many-body interacting dynamics. However, there are noticeable counterexamples to this paradigm provided by two recently investigated single-particle models where a dynamical phase transition occurs without the need of infinitely many degrees of freedom \cite{Angeletti,Nyawo}. 

Here, we consider a complementary setting: we show that it is possible to observe large deviation phase transitions, from a disordered phase to an ordered one, in systems of many, but independent, degrees of freedom. We focus specifically on the case of independently evolving Ising spins. This rather counterintuitive result says that the probability of observing an ordered dynamical fluctuation, in a system evolving in a non-interacting manner, can be surprisingly large. This is a statement about the spontaneous {\em dynamical synchronization} of the independent elementary constituents at the fluctuation level. 

We present examples of these LD transitions by exploring fluctuations in a class of dynamical observables that correspond to time-integrated {\em plaquette} operators. For these, the associated tilted generators that encode their LD behaviour (see below for definitions) have an explicit Kramers-Wannier duality. In the LD formulation, this duality is an exact mapping between weak tilting (corresponding to close to typical behaviour) and strong tilting (corresponding to far from typical behaviour), cf.\ weak-strong coupling in the more standard context of dualities. The presence of the dualities is very informative, as they can help to locate the critical point (at the self-dual point) in the presence of a phase transition, something we exploit in the examples we consider. 

We study two cases in detail. In the first one, the transition to the ordered dynamical phase is continuous and the critical behaviour of the dynamical order parameter is of Ising type. This can be shown analytically as the calculation reduces to solving the transverse field Ising model. In the second example we consider the transition is discontinuous (first-order). Here we can demonstrate the duality analytically, and thus determine the self-dual point at which the transition occurs exactly, but we confirm the existence of a LD transition numerically. We achieve this by devising an efficient trajectory sampling scheme which relies on the non-interacting nature of the dynamics. 

The dynamical phase transitions discussed here seem to point to the existence of a far more general class of trajectory transitions in the large size and long-time arising not due to the complexity of the underlying dynamics but of the properties of the time-integrated observable that is probed.

The paper is organised as follows. In Sect.~\ref{sec:LD} we establish our notation and review the basics of dynamical LD methods. In Sect.~\ref{sec:Models} we define the model and dynamical observables we consider. Sections \ref{TFIM_method} and \ref{TPM_method} presents the main results of the paper. Finally, in Sect.~\ref{sec:Conc} we give our conclusions. In the Appendix we prove the duality of one of the cases we consider.

\section{Stochastic Dynamics and Large Deviations} \label{sec:LD}

We focus on classical stochastic systems with continuous-time Markovian dynamics. Our notation is similar to that of, for example, Refs.~\cite{Lecomte,Budini2011,Oakes,Garrahan2018}. The state of the system is described by a vector $\ket{P_t}$ collecting the probabilities $P_t(C)$ of observing the state in a given configuration $C$ at time $t$
 \begin{equation}
 \ket{P_t}=  \sum_{C} P_t(C)\ket{C}\, .
\end{equation}
In classical stochastic dynamics such a vector evolves according to a master equation (ME)
\begin{equation}
 \partial_t \ket{P_t} =  \mathbb{W} \ket{P_t}\,.
 \label{ME}
\end{equation}
The classical stochastic generator  $\mathbb{W}$ contains information about transition probabilities and the escape rate from a given configuration. Explicitly it reads,
\begin{equation}
 \mathbb{W}=\sum_{C, C'\ne C} W(C \to C')\ket{C'} \bra{C}-\sum_{C} R(C)\ket{C} \bra{C}\,,
 \label{classical_generator}
\end{equation}
where $W(C \to C')$ is the transition rate from $C$ to $C'$ and $ R(C)= \sum_{ C'\ne C}W(C \to C')$ is the escape rate from $C$. 
The dynamics encoded in the ME \eqref{ME} is realised by means of stochastic trajectories. Starting from configuration $C$, the system survives in this state for a random time $\Delta t$, which is distributed exponentially according to $S(\Delta t)=R(C)e^{-R(C)\Delta  t}$, and then jumps into a new configuration according to the rates $W(C\to C')$.
 
As such, a trajectory of total time $t$ consists in a sequence of configurations, $\omega_t=( C_0 \to C_{t_1} \to C_{t_2} \to \dots \to C_{t_n}) $, and waiting times for jumps between them, where
$0 < t_1 < t_2 < t_3 < \dots < t_n < t $  are the times at which the change of configuration occurs, and $C_0$ is the initial configuration. Between the time of the last jump $t_n$, and the final time $t$ of the trajectory, the configuration remains unchanged (i.e., ``survives'' in $C_{t_n}$).
The evolution of the probability state vector $|P_t\rangle $ can be recovered by averaging over all stochastic trajectories. 

To investigate the emergence of ordered dynamical phases and of strong spatio-temporal correlations it is necessary to investigate the full probability of time-integrated observables \cite{TOUCHETTE,touchette2011,Garrahan2018,Jack2019}. We represent observables as vectors, 
\begin{equation}
\langle O|=\sum_C O(C)\langle C|\, ,
\end{equation}
where $O(C)$ is the value of the observable in configuration $C$. Given that classical observables are diagonal in the configuration basis, this is equivalent to representing observables as matrices $\hat{O}$ and multiplying on the left by the {\em flat} state 
\begin{equation}
\label{fs}
\langle - |=\sum_C \langle C| \, , 
\end{equation}
so that $\langle O| = \langle - | \hat{O}$.

We consider dynamical observables defined as time-integral of configuration operators over a trajectory
\begin{equation}
\label{Ot}
{\cal O}[\omega_t]=\int^t_0 dt' \langle O|C_{t'}\rangle \, ,
\end{equation}
where $C_{t'}$ is the configuration of the state at time $t'$ in a stochastic realisation $\omega_t$. Note that while $O$ is a static observable (a function only of the configuration), ${\cal O}$ is a dynamical observable, a functional of the trajectory and extensive time (and in space if $O$ is space-additive). 

For each realisation of the process, i.e.~for each trajectory $\omega_t$, there is associated a value of the time-integrated observable, ${\cal O}[\omega_t]$. Its probability distribution can thus be written as
\begin{equation}
 \pi_t({\cal O})= \sum_{\omega_t} {\cal P}({\omega_t}) \delta\left( {\cal O}[\omega_t] -{\cal O}\right)\,,
\end{equation}
where ${\cal P}({\omega_t})$ is the probability of trajectory $\omega_t$ to occur.
For long times this probability distribution is assumed to obey a large deviation (LD) principle
\cite{TOUCHETTE,touchette2011,Garrahan2018,Jack2019}
\begin{equation}
  \pi_t({\cal O}) \asymp  e^{-t \varphi({\cal O}/t)}\,.
\end{equation}
The so-called LD rate function $\varphi(o)$, which is the analog of an entropy density in the ensemble of trajectories, is a function of the intensive order parameter $o={\cal O}/t$ and gives information about the exponential decay of the probability of observing a large deviation far away from the typical value of $o$. Indeed, this function is positive $\varphi(o)\ge0$, and equal to zero only at the typical outcome of the observable. 

The moment generating function (MGF) for the observable, $Z_t(s)$, defined as 
\begin{equation}
 Z_t(s)= \sum_{{\cal O}} \pi_t({\cal O}) e^{s {\cal O}}  \, ,
 \label{Z}
\end{equation}
also obeys a LD principle for large times
\cite{TOUCHETTE,touchette2011,Garrahan2018,Jack2019}
\begin{equation}
Z_t(s)\asymp e^{t\, \theta(s)} ,
\end{equation}
where $\theta(s)$ is the scaled cumulant generating function (SCGF). Its derivatives evaluated at $s=0$ give the cumulants of $O$ scaled by time. The LD function $\theta(s)$ is the analog of a free energy density for the ensemble of trajectories and $s$ is a parameter conjugate to the observable ${\cal O}$. This parameter has a role akin to the temperature in this dynamical LD setting and it can be seen as a parameter modifying the original probability of the observables. Tuning this parameter allows one to explore the tails of the distribution $\pi_t({\cal O})$.

The  SCGF $\theta(s)$ contains the full statistical information about the observable ${\cal O}$, indeed the $n-$th scaled cumulant of the observable is 
\begin{equation}
 \frac{\langle\!\langle{\cal O}^n\rangle\!\rangle}{t}= \frac{\partial^n}{\partial {s}^n} \theta(s) \bigg\rvert_{s=0}\,,
\end{equation}
where $\langle\!\langle \cdot \rangle\!\rangle$ indicates  cumulant. 
Interesting phase behaviours in trajectory space are encoded in the analytic properties of the SCGF and its singularities are indicative of dynamical phase transitions.
The rate function $\varphi(o)$ and the SCGF $\theta(s)$ are related by a Legendre-Fenchel transform 
\cite{TOUCHETTE,touchette2011,Garrahan2018,Jack2019}
\begin{equation}
 \theta(s)= \sup_{o} \left[ o \, s - \varphi(o) \right] \,. 
  \label{legendre}
\end{equation}
For finite $s$, the SCGF $\theta(s)$ encodes information about large fluctuations  of the stochastic dynamics. More precisely, for a given $s$, $\theta(s)$ provides a way to investigate the trajectories of the process giving an atypical time-average value of $\langle o\rangle_s$ identified by the relation
\begin{equation}
\langle o\rangle_s= \theta'(s)\, .
\label{parameter_Ising_like}
\end{equation}

The SCGF $\theta(s)$ can be obtained by means of tilted operators techniques
\cite{TOUCHETTE,touchette2011,Garrahan2018,Jack2019}.  In particular, it is the largest eigenvalue associated with a deformation of the original stochastic generator $\mathbb{W}$, 
which for observables of the form \eqref{Ot} reads,
\begin{equation}
\begin{aligned}
\label{Wss}
  \mathbb{W}_s&=\sum_{C, C'\ne C} W(C \to C')\ket{C'} \bra{C}-\\
  &\sum_{C} \left[ R(C)-s\,O(C) \right] \ket{C} \bra{C}\,.
\end{aligned}
\end{equation}

\section{Model and Observables} \label{sec:Models}
We now introduce the specific class of models that is addressed in this paper. We consider systems which consist of $N$ non-interacting Ising spins. For each spin, the state space is spanned by two configurations: an up state $\ket{\uparrow}$ indicating the spin has a positive magnetisation and a down state $\ket{\downarrow}$ indicating instead a negative local magnetisation. The magnetisation operator for each spin is the Pauli matrix $\sigma^z$, i.e.~$\sigma^z\ket{\uparrow} =\ket{\uparrow}$  and $\sigma^z\ket{\downarrow} =-\ket{\downarrow}$. The basis state $|C\rangle$ for a configuration $C$  of the whole system can thus be written as a tensor product of the state of each individual spin. 

The probability vector $|P_t\rangle$ evolves according to a non-interacting thermal dynamics
\begin{equation}
 \mathbb{W} = \gamma_{+}\sum_{i=1}^{N} \left[ \sigma_{i}^+ - ({\bf 1}-n_i)\right] +\gamma_{-} \sum_{i=1}^{N} \left( \sigma_{i}^- - n_i\right) \,,
 \label{non_int_dynam}
\end{equation}
where $\sigma_i^{+/-}$ is the ladder operator flipping the $i$-th spin up/down. The operator $n$ is the number operator, defined as $n=\sigma^+\sigma^-$, taking values $0$ or $1$ for a spin down or up, respectively. We denote the rate for the up transition $\ket{\downarrow} \to \ket{\uparrow}$  by $\gamma_+$  and the rate for the down transition $\ket{\uparrow} \to \ket{\downarrow}$ by $\gamma_-$. The shape of the generator, which is a separate sum over spins without any interaction terms, makes explicit the fact that the spins evolve in a completely independent way. The dynamics \eqref{non_int_dynam} is ``thermal'' as it obeys detailed balance. 
The stationary state is therefore an equilibrium state, which given the non-interacting nature of the dynamics takes the form of a product state,
\begin{equation}
\label{eq}
| {\rm eq}_T \rangle = \otimes_{i=1}^N \left( 
\frac{\gamma_+}{\gamma_+ + \gamma_-} \ket{\uparrow}_i + 
\frac{\gamma_-}{\gamma_+ + \gamma_-} \ket{\downarrow}_i \right) \, ,
\end{equation}
where the ``temperature'' is defined in terms of the ratio of the rates, $T = 1/\log{(\gamma_-/\gamma_+)}$ (assuming $\gamma_+ < \gamma_-$). That is, this corresponds to the equilibrium state of a collection of non-interacting spins in a magnetic field of unit strength at temperature $T$.

When $\gamma_+=\gamma_-=\gamma$, up and down transitions have the same probability. In this case, which is the one we consider below, the generator further simplifies to
\begin{equation}
 \mathbb{W} =  \gamma \sum_{i=1}^{N} \left( \sigma_{i}^x - {\bf 1}\right)\,,
 \label{generator}
\end{equation}
where $\sigma^{x}_i=\sigma_i^++\sigma_i^-$ is the first Pauli matrix acting on the $i$-th spin. For this case, the equilibrium state of the system is an equal superposition of all states, i.e.\ the $T=\infty$ state
\begin{equation}
\label{eqint}
| {\rm eq}_\infty \rangle = \otimes_{i=1}^N \left( 
\frac{1}{2} \ket{\uparrow}_i + \frac{1}{2} 
\ket{\downarrow}_i \right) = \, 2^{-N} \, \ket{-} .
\end{equation}
We term the dynamics generated by \eqref{generator} an infinite temperature dynamics. Note that in this case the generator is Hermitian and the left and right eigenvectors coincide (up to normalisation).

\begin{figure}
   \centering
  \includegraphics[scale=0.3]{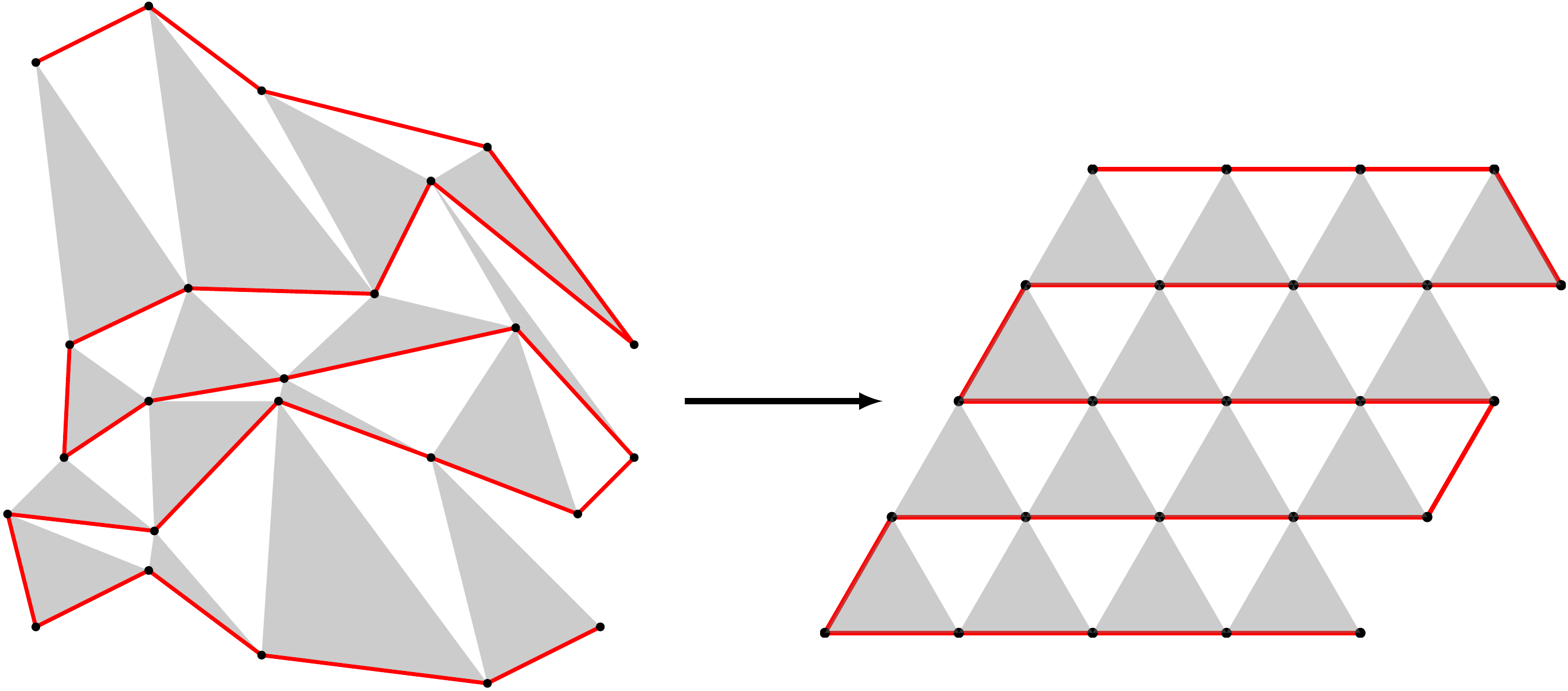}
   \caption{Left: Geometrical illustration of 24 non-interacting spins. We represent them as randomly distributed in space as without interactions there is no notion of a geometrical arrangement. Their dynamics is described by the generator of Eq.~\eqref{generator}. Right: When considering an observable we induce a particular geometry. 
   For Eq.~\eqref{IsingObs} this corresponds to one dimensional sequence, corresponding to the  
  red solid lines connecting the spins. For Eq.~\eqref{TPMObs} it is that a triangular lattice geometry where triplets of spins defining the local plaquettes sit on the vertices of upward-pointing triangles (shaded grey). Here we are representing only a subsection of these lattices therefore we do not describe their corresponding boundary conditions.  \label{TPM_and_Ising_like}}
\end{figure}

For a collection of independently evolving Ising spins, as expected, typical trajectories are uncorrelated and lack of any structure. We are interested in studying large fluctuations of time-integrated observables which represent a measure of ``order" in the spin configuration. 
We will consider what we call in general a {\em plaquette operator} over the spins, of the general form
\begin{equation}
O_{\rm plaquette}(C) = \sum_{\mu} p_\mu(C) \, . 
\label{plaq}
\end{equation}
Here $p_\mu$ are $N$ operators (we always consider periodic boundary conditions) 
\begin{equation}
p_\mu(C) =  \sigma_{i_{\mu_1}}^z \sigma_{i_{\mu_2}}^z \cdots \sigma_{i_{\mu_m}}^z \, ,
\label{pmu}
\end{equation}
formed by the product of $m$ spins. We call the product \eqref{pmu} a plaquette. Each spin in the system belongs to $m$ plaquettes. This means the following: for each configuration of spins $C = \{ \sigma_i^z \}_{i=1}^N$ there is corresponding set of values of the plaquettes $\{ p_\mu \}_{\mu =1}^N$ which is unique (or almost unique, up to symmetries) and therefore a 1-to-1 correspondence between spins and plaquettes. The corresponding dynamical order parameters that we consider then take the form,
\begin{equation}
{\cal O}_{\rm plaquette} = \int_{0}^{t} dt^{\prime} \, p_\mu(C(t')) .
 \label{IsingObs}
\end{equation}

We consider in detail two specific plaquette operators, as they illustrate the range of behaviour that we expect to see more generally. The first one corresponds to $m=2$ and we call it a {\em bond} operator. It is a sum of the simplest plaquettes formed by the product of two spins, 
\begin{equation}
O_{\rm bond} = \sum_{i=1}^{N-1} \sigma_{i}^z \sigma_{i+1}^z + \sigma_{N}^z \sigma_{1}^z \, .
\label{Obond}
\end{equation}
It measures the degree of ordering between consecutive spins. The last term is a ``boundary condition'' connecting the first and last spins. Each term in \eqref{Obond} is a local plaquette $p_\mu$, and the set of them obeys the conditions above: there are $N$ plaquettes; each plaquette is composed of $m=2$ spins and each spin belongs to exactly $m=2$ plaquettes; for each arrangement of the plaquettes there is a unique arrangement of spins, up to a global up-down symmetry in the spins (a consequence of $m$ being even).

The second operator we consider below is the sum of plaquettes with $m=3$ formed by triplets of spins, or {\em triangular plaquettes} 
\begin{equation}
O_{\rm TP} =  \sum_{i,j,k \in \triangle} \sigma_{i}^z \sigma_{j}^z \sigma_{k}^z \, ,
\label{OTP}
\end{equation}
where the sum is over triplets of sites $(i,j,k)$ involved in the triangular plaquette $\triangle$, chosen such that the 1-to-1 condition above is obeyed. Note that by defining the observables we are introducing a certain geometrical arrangement of the otherwise unstructured ensemble of $N$ spins. For $O_{\rm bond}$ the corresponding arrangement is that of a one dimensional chain with periodic boundaries,
where the observable is defined in terms of the bonds between nearest neighbouring spins. For $O_{\rm TP}$, it is that of a two-dimensional triangular lattice, and the plaquettes are upward pointing triangles of nearest neighbours (as in the classical triangular plaquette model, see Refs.~\cite{Newman1999,Garrahan2000,Garrahan2002,Turner}). For $O_{\rm TP}$ we also obey the conditions above: there are as many upward pointing triangles as spins, and since there is no up-down symmetry, plaquette arrangements and spin configurations are 1-to-1 (for periodic boundary conditions in at least one direction, see e.g.~\cite{Garrahan2002}).
The arrangements for the two observables are illustrated in 
Fig.~\ref{TPM_and_Ising_like}.

The corresponding dynamical order parameters are 
\begin{equation}
{\cal O}_{\rm bond}=\int_{0}^{t} dt^{\prime} \sum_{i=1}^N \sigma_{i}^z(t')\sigma_{i+1}^z(t')\,,\label{IsingObs}
\end{equation}
where $i$ runs through the spins and we identify site $N+1$ with site $1$, 
and
\begin{equation}
{\cal O}_{\rm TP}=\int_{0}^{t} dt^{\prime} 
\sum_{i,j,k \in \triangle} \sigma_{i}^z(t')\sigma_{j}^z(t') \sigma_{k}^z(t') \,.
 \label{TPMObs}
\end{equation}
These trajectory observables probe spontaneous fluctuations in the dynamics displaying the two different kinds of order in the spin patterns. 

The associated tilted generators $\mathbb{W}_{s}$, cf.\ \eqref{Wss}, which encode the LD statistics of these observables, are 
\begin{equation}
\label{Tilted_TFIM}
 \mathbb{W}_{s, \text{bond}} = \sum_{i=1}^{N} \left( \sigma_{i}^x - {\bf 1}\right)+s \sum_{i=1}^{N} \sigma_{i}^z \sigma_{i+1}^z\,, 
\end{equation}
and
\begin{equation}
 \mathbb{W}_{s, \text{TP}} = \sum_{i=1}^{N} \left( \sigma_{i}^x - {\bf 1}\right)+s  \sum_{i,j,k \in \triangle} \sigma_{i}^z\sigma_{j}^z\sigma_{k}^z\,, 
 \label{Wstp}
\end{equation}

These two deformed operators look respectively like (up to a sign and an additive constant $N$) the 1D Transverse Field Ising Model (TFIM) with periodic boundary conditions (PBC) \cite{Sachdev,Cabrera}
\begin{equation}
 H_{\text{TFIM}}=-J \sum_{i=1}^N \sigma_{i }^z  \sigma_{i+1 }^z   - h\sum_{i=1}^N \sigma_{i}^x\,,
\end{equation}
and the 2D Quantum Triangular Plaquette Model (QTPM) \cite{Yoshida2014,Turner,Devakul}, \textit{cf}.~Fig.~\ref{TPM_and_Ising_like}
\begin{equation}
 H_{\text{QTPM}}=-J  \sum_{i,j,k \in \triangle} \sigma_{i }^z  \sigma_{j }^z  \sigma_{k}^z - h\sum_{i=1}^N \sigma_{i}^x\, .
\end{equation}
The identification being 
\begin{align}
\label{Wb}
\mathbb{W}_{s, \text{bond}} &= - H_{\text{TFIM}}- N \, , \\
\label{Wtp}
\mathbb{W}_{s, \text{TP}} &= -H_{\text{QTPM}}- N \, , 
\end{align}
with 
\begin{equation}
J=s \;\; \text{and} \;\; h=1 \, .
\end{equation}
These two models have an exact Kramers-Wannier duality. This duality is well known for the TFIM \cite{Sachdev}, and we prove it for the QTPM in the Appendix. Duality symmetries like this one are very informative. They often help to locate the critical point at which a phase transition occurs in the large size limit. For the models above, the self-dual point is at $J=|h|$. This is known to be the phase transition point of the TFIM and, as we show below, also a transition point for the QTPM. As we discuss below, the duality has important consequences for the SCGF $\theta(s)$ and therefore for the statistics of trajectories of the non-interacting spin system.

\section{Large deviations of the bond order parameter}
\label{TFIM_method}

The one dimensional spin-$1/2$ TFIM is one of the most studied models in physics. It allows for amenable analytical considerations and it has direct experimental realisations \cite{Sachdev, Baxter}.
One of its prominent features is an \textit{exact} duality between an ordered and a disordered phase, which has been proved in \cite{Kogut, Kramers, Kramers1}.
Here we  discuss the consequences of this duality in the TFIM with PBC for the SCGF $\theta(s)$.
We denote as $E_0 \left( J, h \right)$ the ground state energy of the TFIM Hamiltonian with PBC $H_{\text{TFIM}} \left( J, h \right)$ \cite{Cabrera,He}, where 
\begin{equation}
\begin{aligned}
 E_0(J, h) = -J  \sum_{\substack{m=0 \\ m\, \text{odd}}}^{2N-1} \left[\left( 1 -\frac{h}{J}\right)^2 + 4\frac{h}{J}\sin^2 \left( \frac{m \pi}{2N}\right)  \right]^{1/2}
 \label{gs_energy_ising}
\end{aligned}
\end{equation}
and perform the duality transformation
\begin{equation}
\begin{aligned}
 H_{\text{TFIM}}(J,h)&=H_{\text{TFIM}}(h, J)\\
 &=J \, H_{\text{TFIM}}( h/J, 1)
\end{aligned}
\end{equation}
which for the ground state energy means
\begin{equation}
E_0( J, h)=  J\, E_0( h/J, 1) \, .
\label{Edual}
\end{equation}

As seen in the previous section, the associated tilted operator describing the statistics of the observable ${\cal O}_{\rm bond}$ is given by \eqref{Wb}. The SCGF is therefore
\begin{equation}
\theta(s) = - E_0(s,1) - N \, ,
\label{thetab}
\end{equation}
and the duality relation \eqref{Edual} gives
\begin{equation}
 \theta(s)= s\,\theta(1/s)+N(s-1)\,.
 \label{theta}
\end{equation}

\begin{figure*}[t]
\includegraphics[scale= 0.46]{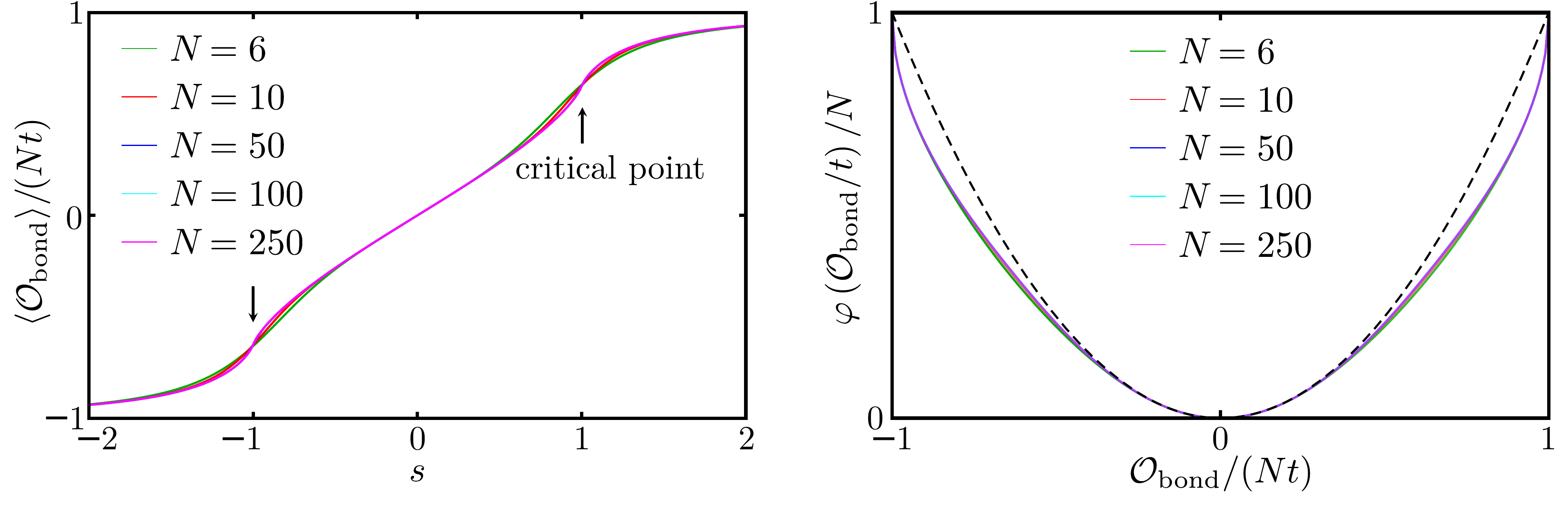}
\caption{Continuous LD transition for bond observable. 
(Left) 
Order parameter $\langle o\rangle_s=\langle {\cal O}_{\rm bond}\rangle /(N t) = \theta'(s)/N$ 
as a function of the counting field $s$, \textit{cf}.~Eqs.~\eqref{parameter_Ising_like} and \eqref{IsingObs}. There is a continuous transition of the 2D Ising type at the self-dual points $s=1$ (to a ferromagnetic phase with finite positive magnetization) and $s=-1$ (to an antiferromagnetic phase with finite negative magnetization). 
Around $s=0$  we observe a linear trend in $\langle o\rangle_s$, while for $s>1$ ($s<1$) it saturates asymptotically to its minimum and maximum values. Here we display $\langle o\rangle_s$ with appropriate scaling for different system sizes by means of Eq.\eqref{gs_energy_ising}.
(Right)
Rate function $\varphi(o)$ for the observable ${\cal O}_{\rm bond}$ \eqref{IsingObs} for different system sizes. We compare this function with the Gaussian LD rate function accounting for the Gaussian fluctuations of the noninteracting spin system. This has zero mean and variance $\sigma^2=\theta''(0)=0.5$ (back dashed curve). The existence of a singularity of this observable is manifested in the broadening of $\varphi(o)$ with respect to the Gaussian distribution, indicating that fluctuations of the order parameter are correlated.}
\label{Obs_TFIM}
\end{figure*}

If we take the derivative with respect to $s$ of \eqref{thetab} we get the average of the order parameter, $\langle {\cal O}_{\rm bond} \rangle / t$, \textit{cf}.~Eq.\eqref{IsingObs}. In Fig.~\ref{Obs_TFIM}(left) we plot it for a range of system sizes. We observe a second order phase transition at the self dual points, which are $s=1$ and, because of a $\mathbb{Z}_2$ symmetry, also $s=-1$. Finite size effects are present for the smaller sizes shown, $N=6,10$, but for $N=50$ and beyond the SCGFs show convergence, with the exception of points close to the transition, as expected.

Using the relation between the SCGF and the rate function as defined by Eq.~\eqref{legendre} we plot the rate function in Fig.~\ref{Obs_TFIM}(right). The transitions that can be seen in the observables appear in the rate function as a broadening of the rate function. This broadening is due to the transition between two different dynamical phases of the model. 
The rate function being broader than the corresponding Gaussian (dashed line), with variance given by the second scaled cumulant of the observable, shows that the fluctuations associated with the order detected by \eqref{IsingObs} are more pronounced that one could have anticipated from the non-interacting nature of the dynamics.

\section{Large deviations of the triangular plaquette order parameter}
\label{TPM_method}

\subsection{Sampling LDs with transition path sampling}

The same relation as in Eq.~\eqref{theta} holds also for the SCGF of the TPM, but unlike the TFIM, this model is not exactly solvable. In order to numerically access the statistics of the triangular plaquette observable \eqref{TPMObs} in the trajectories of the non-interacting spin system  we employ the method of transition path sampling (TPS) \cite{Bolhuis}, as adapted for the study of LDs, see e.g.\ \cite{Hedges2009,Oakes,Mora}. 
The basic idea behind TPS is similar to that of Markov chain Monte Carlo but applied to trajectories rather than configurations \cite{Bolhuis}. For the case of LDs, it amounts to an importance sampling method that helps overcome some of the exponential (in size and time) cost of sampling the  rare event. TPS is well suited for dynamics that is reversible - like the infinite temperature dynamics we study here - for reasons we explain below. Other numerical methods for LDs include sampling exponentially rare trajectories via ``cloning'' (also known as splitting) \cite{Giardina2011}, variational approximations to rate functions \cite{Jacobson2019}, or tensor network methods to directly estimate SCGFs \cite{Banuls2019,Helms2019}.  

TPS \cite{Bolhuis,Hedges2009,Oakes} does a biased random walk in trajectory space (just like normal Monte Carlo does in configuration space) by starting from a seed trajectory and updating it by proposing modifications to it. These are accepted or rejected according a Metropolis rule, that is, with probability of acceptance $\min{\left( 1, e^{s \Delta {\cal O}} \right)}$, where $\Delta {\cal O}$ is the change in the observable of interest in the case of LD studies, cf.\ Eq.~\eqref{Z}. A key aspect of TPS is how to propose trajectories. In two commonly used approaches, such as ``shooting'' or ``shifting'' \cite{Bolhuis}, the proposed update to the trial trajectory begins by selecting a given time $\tau_{\textnormal{cut}}$ during the previous trajectory. Starting from the configuration at $\tau_{\textnormal{cut}}$ in the old trajectory, a new section of trajectory is produced ending at either the end of the trajectory $t_{\textnormal{max}}=t$ (in the case of forward dynamics) or at the start of the trajectory $t=0$ (in the case of reversible dynamics). In either case, the new section of trajectory is bounded only by configuration at $\tau_{\textnormal{cut}}$, since the configurations at either $t=0$ or $t_{\textnormal{max}}$ can be changed. This is a desired property of TPS methods, since the initial and final configurations will typically be sampled using the original dynamics, i.e.\ they will be representative of the $s=0$ distribution of trajectories. 

The drawback of the above ways of proposing trajectories comes in the convergence (in terms of TPS iterations) of the bulk of the trajectory. Since the acceptance rule is exponential in the change $\Delta {\cal O}$, smaller updates are accepted exponentially (in time) more frequently. The consequence is that the ends of the trajectories converge more quickly that its bulk, which gets updated very rarely. This means that the decorrelation of TPS moves takes a number of iterations that increases exponentially both in time and system size.

\subsection{Improvement to TPS trajectory proposal and acceptance} 

To sample the desired trajectory ensemble $\langle O(\omega) e^{-s O(\omega)} \rangle$ we use a modified form of the TPS method that takes advantage of the non-interacting nature of our spin dynamics. We do two things that allow to overcome the exponential in time and size cost of TPS equilibration.

First, since the dynamics is non-interacting we can propose a new trajectory from an old one by {\em only changing the dynamics of a single spin}. This guarantees that the change 
$\Delta {\cal O}$ will be of order 1 in terms of system size, rather than order $N$ as usual (while still for the moment extensive in time). This obvious trick already represents a significant speed up to the acceptance of TPS moves. Of course, this is only possible in a non-interacting case as the one we are considering. 

The second improvement is to propose only changes in a single spin trajectory which are {\em localised in time}. Specifically, rather than using $\tau_{\textnormal{cut}}$ as the only point along the spin trajectory defining the change, we also introduce a window of variable time for the new section of trajectory, whose maximum is $\tau_{\textnormal{window}} = \tau_{\textnormal{cut}} + \tau_{\textnormal{lim}}$, where $\tau_{\textnormal{lim}}$ is uniformly sampled from $1 \leq  \tau_{\textnormal{lim}} \leq 0.2\, \tau_{\textnormal{max}}$.
The boundary conditions imposed at the two edges of the window are dictated by the orientation of the spins at the times $\tau_{\textnormal{cut}}$ and $\tau_{\textnormal{window}}$. That is, the orientation of the spin in the new trajectory at $\tau_{\textnormal{cut}} ( \tau_{\textnormal{window}})$ has to match up with the orientation that it had at $\tau_{\textnormal{cut}} ( \tau_{\textnormal{window}})$ in the old trajectory. Since each spin can take only one of two values, this limitation does not impose too much of an extra computational cost. This cost of discarding trajectories that do not satisfy the boundary conditions of the time window is compensated by the ability of these windows to equilibrate (in the TPS sense) the whole trajectory faster. The reason is that between the two improvements $\Delta {\cal O}$ becomes order $1$ in both the  space and time extent. 

\begin{figure}
   \centering
  \includegraphics[scale=0.45]{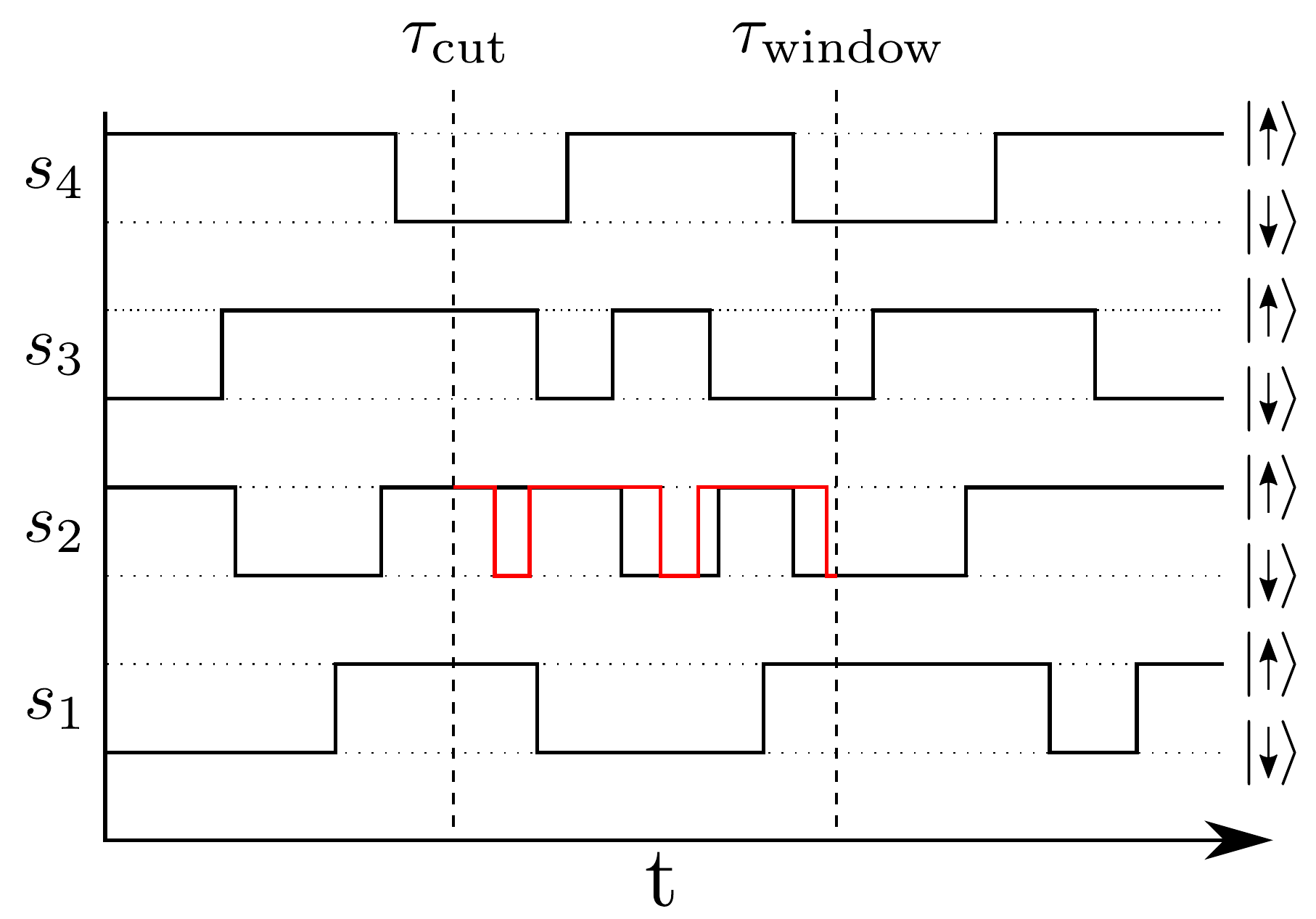}
   \caption{WTPS trajectory proposal scheme. This schematic shows a system of four spins $s_i$, where $i=1,2,3,4$. The many-body trajectory is composed of four individual spin trajectories which are independent due to the non-interacting underlying dynamics (black solid lines indicate the original trajectory). Each spin is either up or down. The proposed trajectory changes a single spin only, $i=2$ in the sketch.  The change starts at time $\tau_{\textnormal{cut}}$
   and ends at $\tau_{\textnormal{window}}$ as described in the main text. Only for the selected spin a new part of the trajectory is created between $\tau_{\textnormal{cut}}$ and $\tau_{\textnormal{window}}$ (red solid line). \label{WTPS}}
\end{figure}

Fig.~\ref{WTPS} shows a schematic of how a new trajectory is produced by modifying only the chosen window $\tau_{\textnormal{window}} $ of the selected spin trajectory rather than the usual update to part of the trajectory for the whole system as used in other typical TPS methods \cite{Bolhuis}.

\subsection{Trajectory umbrella sampling and choice of alternative dynamics}

In order to efficiently sample trajectories for value of $s$ across the LD transition, expected to occur at the self-dual points $s = \pm 1$, we need further enhancements to the efficiency of the TPS scheme. We achieve this by means of umbrella sampling in trajectory space \cite{Ray,Ray1,Nemoto,Klymko,Oakes}. 

We are trying to calculate quantities like
\begin{equation}
\label{Aexp}
\langle {\cal A}(\omega) e^{s {\cal O}(\omega)} \rangle
= \sum_{\omega} {\cal P}(\omega) \, {\cal A}(\omega) \, e^{s {\cal O}(\omega)} 
\, ,
\end{equation}
by sampling trajectories in the exponential tilted ensemble, where TPS deals with the exponential  factor in the expression above. The idea of umbrella sampling is to exploit the identity,
\begin{align}
\langle {\cal A}(\omega) e^{s {\cal O}(\omega)} \rangle
&= \sum_{\omega} {\cal P}(\omega) \, {\cal A}(\omega) \, e^{s {\cal O}(\omega)}
\nonumber \\
&= \sum_{\omega} {\cal P}_{\rm ref}(\omega) \, 
\frac{{\cal P}(\omega)}{{\cal P}_{\rm ref}(\omega)} \,
{\cal A}(\omega) \, e^{s {\cal O}(\omega)} 
\nonumber \\
&= 
\langle {\cal A}(\omega) e^{s {\cal O}(\omega)} e^{{\cal G}(\omega)} \rangle_{\rm ref} \, ,
\label{us}
\end{align}
where 
\begin{equation}
{\cal G}(\omega) = \log {\cal P}(\omega) - \log {\cal P}_{\rm ref}(\omega) \, .
\end{equation}
Equation \eqref{us} means that we can estimate \eqref{Aexp} by means of a ``reference'' dynamics different from the original dynamics and adjusting through the exponential of ${\cal G}$ to account for the change in measure over the trajectories \cite{Ray,Ray1,Nemoto,Klymko,Oakes}.  
Chosen judiciously, the reference dynamics may reduce the sampling error due to the exponential weighting in \eqref{us}. There is an optimal choice for this reference, that of the dynamics obtained via a generalised Doob transformation \cite{Jack,Chetrite2015,Carollo2018}. This optimal choice is in general impossible to implement explicitly (as it requires the diagonalisation of the tilted generator) and one has to resort to tractable approximations.  

\begin{figure*}[t]
\includegraphics[scale= 0.46]{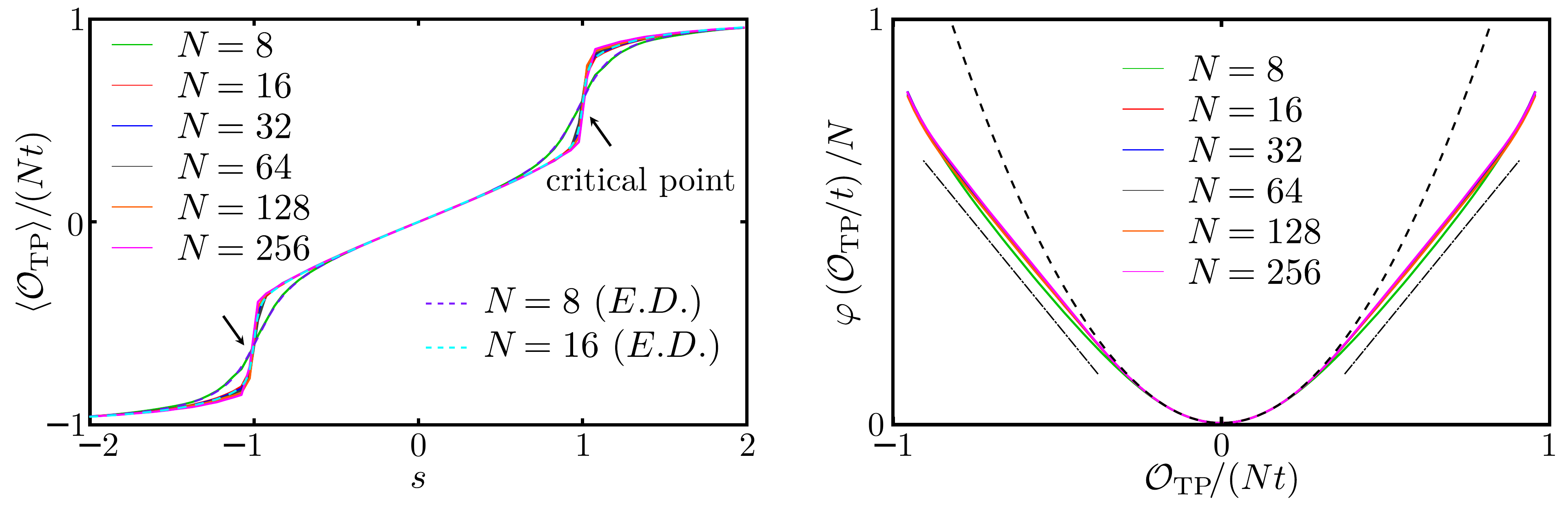}
\caption{
LD transition for the triangular plaquette observable. 
(Left) 
Discontinuous phase transition of time-averaged value of the TPM-like plaquette observable at $s=1$ and $s=-1$. Data obtained by utilizing the method discussed in Sec. \ref{TPM_method} are shown for various system sizes. As a further numerical tool to improve the analysis of the data we used multistate Bennett acceptance ratio (MBAR) \cite{ Shirts,BENNETT}. Also for this model we observe linear regime around $s=0$ and asymptotic behavior beyond the critical point $s>1$ and $s<-1$. 
(Right) Rate function $\varphi(o)$ for the observable ${\cal O}_{\rm TP}$ \eqref{TPMObs} for different system sizes. This is also compared with the Gaussian fluctuations of the noninteracting spin system. In this case such Gaussian LD rate function has zero mean and variance $\sigma^2=\theta''(0)\approx 0.34$ (back dashed curve). The existence of a first-order phase transition is manifest in the linear behaviour of $\varphi(o)$ in the region delimited by the values across the first-order jumps in the SCGF of the left panel. The rate function is thus much broader than the corresponding Gaussian, indicating the higher likelihood of trajectories with plaquette order. The slope of the linear parts is given by the critical values $s=\pm 1$, shown as the dot-dashed lines for comparison. 
}
\label{Obs_TPM}
\end{figure*}

For the specific model we are investigating, we choose dynamics to maintain the interesting nature of a collective phase transition for a set of stochastic independently evolving spins. The plaquette observable we are considering, cf.~\eqref{Obs_TPM}, may suggest a reference dynamics that contains interactions. But to really emphasise that this collective behaviour emerges from the non-interacting nature, we chose a reference dynamics which is also non-interacting. 

Specifically, we choose dynamics generated by \eqref{non_int_dynam}, where the reference dynamics differs from the original one in that $\gamma_{+} \neq \gamma_{-}$. That is, we use ``temperature'' as the control parameters of the trajectory umbrella sampling. The exponent ${\cal G}$ of the reweighing factor is easy to calculate in terms of the original dynamics rate, $\gamma$, and those of the reference, $\gamma_\pm$. In order to sample \eqref{us} we ran TPS with the reference dynamics, under the proposal rules described in the previous subsection, and with acceptance criterion where $s \Delta {\cal O}$ is replaced by $s \Delta {\cal O} + \Delta {\cal G}$ \cite{Ray,Ray1,Nemoto,Klymko,Oakes}.

\subsection{First order transition in triangular plaquette order}

In Fig.~\ref{Obs_TPM}(left) we show the results of sampling the average of the observable \eqref{TPMObs} using the TPS scheme described above 
for system sizes $N = 8,16,32,64,128,256$ (full lines). For comparison we also show the results from exact diagonalisation (E.D.) of small sizes, $L=8, 16$ (dashed lines). 
For increased system size it is clear that the order parameter tends towards becoming discontinuous at the self-dual points $s=1$ and $s=-1$. This indicates that the trajectory transition is of the first-order kind, between a dynamically disordered phase for $|s| < 1$, and dynamical phases with triangular plaquette order for $|s|>1$. 

In Fig.~\ref{Obs_TPM}(right) we show the corresponding rate function. We observe that the latter has broader tails than as compared to a Gaussian rate function (i.e., a parabola) with the same average and variance. We see that the first-order nature of the transition is reflected in a linear behaviour of the rate function for values of the observable which lie in between the extreme values across the discontinuous jumps of the SCGF at $s=\pm1$. The linear behaviour of the rate function correspond to a Maxwell construct for the first-order transitions of the SCGF. This becomes exact in the $N \to \infty$ limit, and the slope which coincides with the critical $s$-field values ($s=-1$ and $s=1$), shown for comparison in Fig.~\ref{Obs_TPM}(right) as the dot-dashed lines. The broadness of the rate function indicates the enhanced probability of the triangular plaquette order in the dynamics.

Operators similar to \eqref{Wstp} have been considered before in the context of quantum codes \cite{Yoshida2014} and of ``fractons'' \cite{Devakul}. The first order transition we find numerically at the self-dual point via our augmented TPS scheme is also compatible with the results of Ref.\ \cite{Yoshida2014}.

\section{Conclusions} \label{sec:Conc}

In this paper we have investigated the dynamical fluctuations of a collection of independent spins evolving stochastically. We have shown that despite the non-interacting nature of their dynamics, the statistics time-integrated interacting observables can be non trivial, and give rise to correlated behaviour at the fluctuation level. We considered two examples of a general class of plaquette observables, showing that in the large number of spins limit they give rise to both continuous and first order LD transitions. 

Our findings here fit with recent results for few body problems \cite{Angeletti,Nyawo} where singularities in long-time trajectory ensembles were not a consequence of interactions in the dynamics in the large size limit (like in most other systems displaying LD transitions \cite{Garrahan2018,Jack2019}) but of the properties of the dynamical observables probed. We focused on spin systems and observables where tilted generators displayed a duality symmetry, which we exploited to identify precisely the location of the dynamical transitions. Our approach here directly generalises to other plaquette observables for which their tilted generators will have dualities. These cases will display similar dynamical transitions at their self-dual points. 

We expect the kind of behaviour we uncovered in this paper to be more widespread. Our results suggest that a large collection of non-interacting degrees of freedom can have atypical fluctuations where the system dynamically synchronises and behaves cooperatively for long times. Furthermore, such rare events would correspond to dynamical singularities in the ensemble of trajectories, making them much more likely than the non-interacting nature of the dynamics would indicate. It will be interesting to explore this spontaneous synchronisation at the fluctuation level beyond the simple models considered here.

\acknowledgments
The research leading to these results has received funding from a VC Scholarship for Research Excellence (LMV) and an EPSRC Doctoral Prize (THEO) from the University of Nottingham, and from EPSRC Grant No. EP/N03404X/1 (FC, JPG). We are grateful for access to the University of Nottingham's Augusta HPC service. We also acknowledge the use of Athena at HPC Midlands Plus.

\appendix*

\section{Duality of the Quantum Triangular Plaquette Model} \label{sec:Dual}

In this appendix we prove the duality of the Quantum Triangular Plaquette Model (QTPM). 
To do this we make use of the generalization in higher dimensions of the Kramers-Wannier duality \cite{Kramers} of the classical Ising model in $2D$ \cite{XuMoore}.
This method uses the mapping between the $D$-dimensional classical statistical mechanical models and 
the $d$-dimensional quantum Ising models where $D=d+1, D>1$ \cite{Sachdev}; this technique allows us to show the duality of the QTPM in its classical form for simplicity.
\begin{figure}[t]
\includegraphics[scale= 0.6]{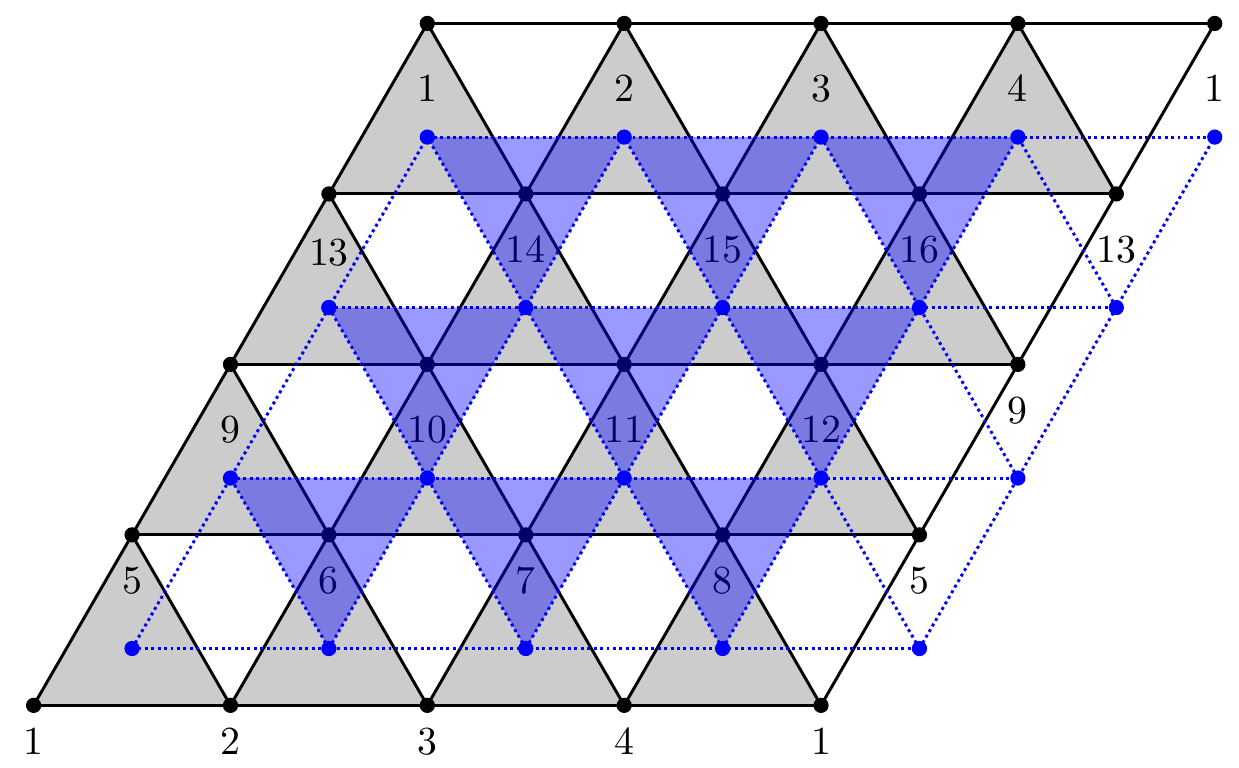}
\caption{ Geometrical illustration of the QTPM with periodic boundary conditions for 16 spins (grey triangles). The spins $\sigma^{\beta}_i$ are positioned at the vertices of upward-pointing triangles $\triangle$ of the original lattice. In the dual problem (blue triangles), the dual spins $\tau_i^{\beta}$ are located at the vertices of downward-pointing triangles  $\triangledown$, which are at the centers of the original plaquettes, on the dual lattice.}
\label{dualita_reticolo_triangolare}
\end{figure}

\begin{figure}[h]
   \centering
  \includegraphics[scale=0.7]{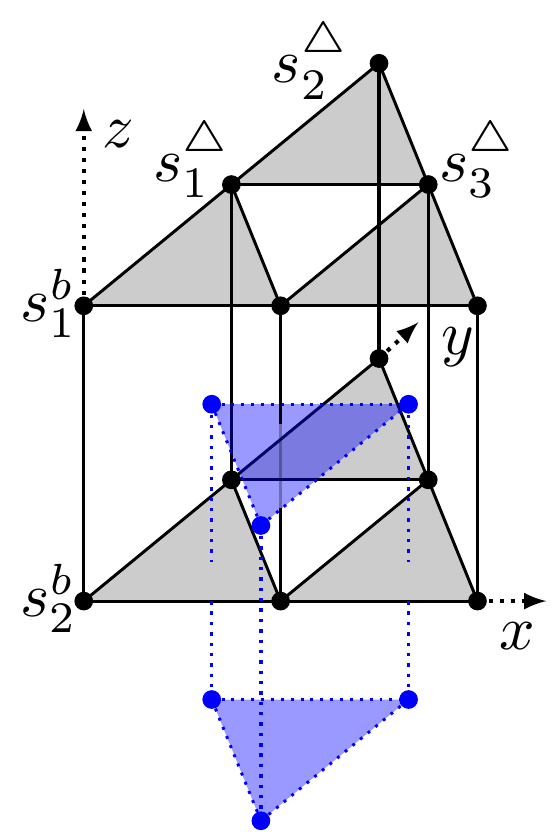}
   \caption{Geometrical illustration of plaquette $\triangle = s_1^{\triangle} s_2^{\triangle} s_3^{\triangle}$ and bond $ b= s_1^{b} s_2^{b}$ interactions on the original lattice (gray triangles) in the classical $3D$ model. 
  In the dual problem (blue triangles), the spins are located on the dual lattice at the vertices of downward-pointing triangles $\triangledown$. The blue prism is the geometrical illustration 
  of the plaquette and bond interactions on the dual lattice in the classical $3D$ model. \label{dualita_triangolare_3d}}
\end{figure}

The $2d$ QTPM is described by a plaquette spin model, defined in terms of Ising spins, on a regular triangular lattice, with Hamiltonian
\begin{equation}
 H_{\text{QTPM}}=-J \sum_{\triangle} \sigma_{1 }^z  \sigma_{2 }^z  \sigma_{3}^z - h\sum_{i} \sigma_{i}^x\,,
\end{equation}
where $\sigma^{\beta}_i$, with $\beta=x,y,z$, are the local Pauli matrices in the direction $\beta$ acting on the $i-$th site. 
The interaction, embodied in the parameter $J$, occurs between triplets of spins, which for simplicity we denote as $\sigma_{1 }^z, \sigma_{2 }^z,\sigma_{3 }^z $, positioned at the vertices of upward-pointing triangles $\triangle$,
while $h$ is a transverse magnetic field in the $x$ direction.
We consider a system whose  linear size  is an integer power of two and periodic boundary conditions \cite{Martin, Juan1, Turner,Juan2}, \textit{cf}.~Fig.~\ref{dualita_reticolo_triangolare}.  \\ 
The energy function of the corresponding $3D$ classical model has the form
\begin{equation}
 \label{eq:2}
 \beta E =-K \sum_{\triangle} s_{1}^{\triangle}  s_{2}^{\triangle}  s_{3}^{\triangle} - J_z\sum_{b} s_{1}^b  s_{2}^b\,,
\end{equation}
where $s^{\alpha}_{i}=\pm 1$, with $\alpha=\triangle, b$ are Ising variables corresponding to the $i-$th site.
The first sum in the right-hand side of~\eqref{eq:2} is over all triangular plaquettes $\triangle$ in the $xy$ plane, where $K$ represents 
the magnetic interaction among the dynamic variables $s^{\triangle}_{i}$, 
while the second sum is over all the bonds $b$ in the $z$ direction with coupling constant $J_z$, \textit{cf}.~Fig.~\ref{dualita_triangolare_3d}. 
The mapping relations between classical and quantum coupling constants are the conventional ones \cite{Sachdev, XuMoore}:
$K=a J$, $e^{-2 J_z}=\tanh\left( ah \right) $ and  $T=1/(Ma)$, where $a$ is the lattice spacing, $M$ is the number of sites in the $z$ direction of the classical model, and
$T$ is the temperature in the quantum model. 
The mapping between the quantum and the classical model becomes exact
in the scaling limits:  $ a \rightarrow 0$, $ M \rightarrow \infty$ and $T \rightarrow 0$\,.
The partition function of the $3D$ classical model is
\begin{equation}
\begin{aligned}
 Z&= \sum_{\{s\}} e^{-\beta E} \\
 &= \sum_{\{s\}} e^{K \sum_{\triangle} s_{1}^{\triangle}  s_{2}^{\triangle}  s_{3}^{\triangle} + J_z\sum_{b} s_{1}^b  s_{2}^b }\\
 &=  \sum_{\{s\}} \Bigg[ \prod_{\triangle} \left( \cosh K + s_{1}^{\triangle}  s_{2}^{\triangle}  s_{3}^{\triangle} \sinh K \right) \\
 &   \times \prod_{b} \left( \cosh J_z + s_{1}^{b}  s_{2}^{b}  \sinh J_z \right)\Bigg]\,.  
\end{aligned}
\end{equation}
By defining the face variables $k_{\triangle}=0,1$\,, the bond variables $k_{b}=0,1$  and the constants
$c_0=\cosh K$, 
$c_1=\sinh K$, 
$d_0=\cosh J_z$ and
$d_1=\sinh J_z$, we can express the partition function as
\begin{equation}
\begin{aligned}
 Z&=\sum_{\{s\}} \sum_{k_{\triangle}} \sum_{k_b} \Bigg\{ \Bigg[ \prod_{\triangle} c_{k_{\triangle}} \left( s_{1}^{\triangle}  s_{2}^{\triangle}  s_{3}^{\triangle} \right)^{k_{\triangle}} \Bigg] \\
 &\times \left[ \prod_{b} d_{k_{b}} \left( s_{1}^{b}  s_{2}^{b} \right)^{k_{b}} \right]\Bigg\}\,.
\end{aligned}
\end{equation}
We can note that if an Ising variable $s^{\alpha}_{i}$  is raised to an even power then we obtain a factor $2$ in the partition function $Z$, 
originated from the sum over all the spin configurations, otherwise  we get a factor $0$.
This can be expressed in formulae by rewriting the partition function $Z$ as a constrained sum $\sum^{'}$ over the $k$ variables
\begin{equation}
 Z=2^N {\sum_{k_{\triangle}, k_b}}^{'} \Biggl( \prod_{\triangle} c_{k_{\triangle}} \Biggr) \left(  \prod_{b} d_{k_{b}} \right), 
\end{equation}
where  $N$ is the total number of sites. 
Since every site of the original lattice belongs to $3$ triangular face terms and to $2$ bond terms, 
the restriction on the partition function is equivalent to ask that the sum of all $5$ $k_{\triangle /b}$ variables has to be an even number for each site.
In order to solve this constraint dual variables are introduced. 
The dual spins ${s}^*$ are positioned at the centers of the triangular prisms of the original prismatic lattice, \textit{cf}.~Fig.~\ref{dualita_triangolare_3d}. 
For each site $i$, belonging to the original lattice, $3$ vertical bonds, of the dual lattice, bisect the $3$ neighboring spacelike faces $\triangle$ of the site $i$.
Thus, each bisecting bond $b$ of the dual lattice is given by the relation $k_{\triangle}= \frac{1}{2} \left( 1- {s^*}_{1}^b {s^*}_{2}^b  \right)$.
Each spacelike face $\triangledown$ of the dual lattice is pierced by a vertical bond $b$ between two spins of the original lattice, 
thus the plaquette on the dual lattice is set by
$ k_{b}= \frac{1}{2} \left( 1- {s^*}_{1}^{\triangledown} {s^*}_{2}^{\triangledown} {s^*}_{3}^{\triangledown} \right)$, \textit{cf}.~Fig.~\ref{dualita_triangolare_3d}.
The constraint on the variables $k_{\triangle /b}$ is fulfilled since the $6$ dual lattice sites around one site on the original lattice meet the condition
\begin{equation}
 \begin{split}
   k_{\triangle_{1}}  +& k_{\triangle_{2}}+ k_{\triangle_{3}}+ k_{b_1}+ k_{b_2}=\\
& \frac{1}{2} \left( 1- {s^*}_{1}^b {s^*}_{4}^b \right)+ \frac{1}{2} \left( 1- {s^*}_{2}^b{s^*}_{5}^b \right)+ \\
 & \frac{1}{2} \left( 1- {s^*}_{3}^b {s^*}_{6}^b \right)+ \frac{1}{2} \left( 1- {s^*}_{1}^{\triangledown} {s^*}_{2}^{\triangledown} {s^*}_{3}^{\triangledown} \right)\\
& +\frac{1}{2} \left( 1- {s^*}_{4}^{\triangledown} {s^*}_{5}^{\triangledown} {s^*}_{6}^{\triangledown} \right)\equiv 0 \; (\text{mod}\, 2).
 \end{split}
\end{equation}
In order to calculate the dual constant couplings $(\tilde{K},\tilde{J}_z)$ it is useful to rewrite $c_k=k \sinh K+ \left(1-k \right) \cosh K$ by using the fact that 
$c_k$ has been defined for a face $\triangle$ of the original lattice that is pierced by a Ising bond $b$ in the dual problem
\begin{equation}
\begin{split}
 c_{k_{\triangle}}&= \frac{1+ {s^*}_{1}^b {s^*}_{2}^b }{2} \cosh K + \frac{1- {s^*}_{1}^b {s^*}_{2}^b }{2} \sinh K \\
 &= \frac{1}{2}e^{K} \left( 1+ {s^*}_{1}^b {s^*}_{2}^b \; e^{-2 K} \right)\\
 &= \frac{1}{\sqrt{2 \sinh 2  \tilde{J}_z}} e^{\tilde{J}_z {s^*}_{1}^b {s^*}_{2}^b }, \label{eq1}
 \end{split}
\end{equation}
with $\tilde{J}_z$ determined by $e^{-2 K}=\tanh \tilde{J}_z$.
By defining $\tanh \tilde{K}= e^{-2 J_z}$ we obtain
\begin{equation}
\begin{split}
 d_{k_b}&=k_b \sinh J_z+ \left(1-k_b \right) \cosh J_z\\
&= \frac{1}{\sqrt{2 \sinh{2 \tilde{K} }}} e^{\tilde{K} {s^*}_{1}^{\triangledown} {s^*}_{2}^{\triangledown} {s^*}_{3}^{\triangledown} }. \label{eq2}
\end{split}
\end{equation}
Using the results of equations \eqref{eq1} and \eqref{eq2}, we can write, for the $3D$ classical model, 
the relation between the partition function of the original lattice $Z ( K, J_z )$ and dual lattice $Z ( \tilde{K}, \tilde{J}_z )$ 
\begin{equation}
\begin{aligned}
 &Z ( K, J_z )= \\
 &\frac{1} {\text{Vol}(G)} \left(\sinh2\tilde{J}_z \right)^{-\frac{N}{2}}
\left(\sinh 2  \tilde{K} \right)^{-\frac{N}{2}} Z ( \tilde{K}, \tilde{J}_z ),
\end{aligned}
\end{equation}
and the duality relation for the coupling constants
\begin{equation}
\begin{aligned}
 \sinh 2 J_z \sinh 2 K&=1,\\
\end{aligned}
\end{equation}
where the size of the gauge group $\text{Vol}(G)=1$ because if the linear size of the system is an integer power of two and the system has periodic boundary conditions in at least one direction of the lattice,
the ground state is unique, 
thus one dual spin configuration corresponds to one spin configuration in the original lattice \cite{Juan1}. \\
The corresponding $2d$ quantum model has the same duality 
\begin{equation}
\begin{aligned}
  H_{\text{QTPM}}&=-J \sum_{\triangle} \sigma_{1 }^z  \sigma_{2 }^z  \sigma_{3}^z - h\sum_{i} \sigma_{i}^x \\
 &=- J\sum_{i} \tau_{i}^x -h \sum_{\triangledown} \tau_{1 }^z \tau_{2 }^z  \tau_{3}^z,\\ \quad
& \sinh 2 J \sinh 2 h=1, 
\end{aligned}
\end{equation}
where $\tau_i^{\beta}$, with $\beta=x,y,z$, are the dual Ising spins.\\
This Kramers-Wannier duality shows that if the QTPM has a phase transition at zero temperature then the transition should take place at the self dual point $J/h=1$.

\vspace{4cm}

\bibliography{LD}
\bibliographystyle{apsrev4-1} 
\end{document}